\documentclass[12pt,preprint]{aastex}

\usepackage{amsmath}

\begin{document}



\def\lsun{{\rm L_{\odot}}}
\def\msun{{\rm M_{\odot}}}
\def\rsun{{\rm R_{\odot}}}
\def\lta{\la}
\def\gta{\ga}
\def\be{\begin{equation}}
\def\ee{\end{equation}}
\def\lsun{{\rm L_{\odot}}}
\def\le{{L_{\rm Edd}}}
\def\msun{{\rm M_{\odot}}}
\def\rsun{{\rm R_{\odot}}}
\def\rp{{R_{\rm ph}}}
\def\rs{{R_{\rm s}}}
\def\mo{{\dot M_{\rm out}}}
\def\me{{\dot M_{\rm Edd}}}
\def\tc{{t_{\rm C}}}
\parindent0pt


\title{Accretion Disc Theory since Shakura and Sunyaev}
\author{ Andrew~King\altaffilmark{1}}

\altaffiltext{1}{Theoretical Astrophysics Group,
University of
Leicester, Leicester LE1 7RH, U.K.; ark@astro.le.ac.uk}

\begin{abstract}
\noindent 
I briefly review the progress of accretion disc theory since
the seminal paper of Shakura and Sunyaev.
\end{abstract}

\section{Introduction}
Discs are a natural occurrence in astrophysical systems whenever 
they have significant angular momentum. Astronomers discussed
discs of various kinds for much of the twentieth century. However
the paper of Shakura and Sunyaev (1973) transformed the
subject, partly because it unified concepts already discussed,
and partly through technical innovation.

The most important results of the paper were

(a) the condition for an an accretion disc to be {\it thin}, i.e.  to
have scaleheight $H$ much less than disc radius $R$: Shakura \&
Sunyaev (1973) showed that the conditions `thin', `efficiently cooled'
and `Keplerian' are precisely equivalent: if one of them fails, so do
the other two.

(b) it is perfectly possible for an accreting object to be
supplied with mass at a rate that would ultimately produce 
a luminosity above the Eddington limit. Shakura \& Sunyaev (1973) suggested 
that much of the excess would be blown away by radiation
pressure at the radius where this luminosity was first reached, and
the remainder at smaller radii. 
We now know that this is probably what occurs in SS433, and most, if 
not all, ultraluminous X--ray sources (ULXs) (cf Begelman et al., 2006;
Poutanen et al., 2007).

(c) the effective temperature profile of a steady thin disc goes as $T(R)
\propto R^{-3/4}$, and this result is {\it independent} of the
mechanism making gas lose angular momentum and spiral inwards.
This effective temperature profile 
is now well attested, particularly by observations of CVs.
The overall stretched--out blackbody--like continuum spectrum
agrees with it, and more directly, the surface brightness
distribution measured during accretion disc eclipses in CVs
also agrees: eclipses are broad and shallow at long wavelengths
and deep and narrow at short ones.

(d) the mechanism for angular momentum removal
 (Shakura \& Sunyaev (1973) called it `viscosity')  may be 
magnetic in origin. We shall see that this too was a prescient suggestion.

(e) one can parametrize the (kinematic) viscosity
as $\nu = \alpha c_sH$, where $c_s$ is the local sound speed, and
$\alpha$ a quantity of order unity. 
This `alpha prescription' has the great virtue of neatly separating
the `vertical' and `horizontal' structure of a thin disc. Moreover
many quantities of physical interest turn out to depend only weakly on
$\alpha$, suggesting that one can make some progress without knowing
its origin. However it
is vital to realise that {\it accretion disc theory is still
  incomplete}, since we do not know the full spatial and temporal
dependence $\alpha({\bf x}, t)$. 
Viscosity plays a similar role in accretion disc theory to that played
by nuclear burning in stellar evolution theory in the early 20th
century.

\section{Progress}

Just as astronomers were nevertheless able to make some
progress with stellar structure theory,
despite not understanding nuclear burning (cf Eddington's book
{\it The Internal Constitution of the Stars}), theorists and observers
have managed to understand how accretion
discs behave in some situations. Much of this understanding
has come because the disc diffusion equation
\begin{equation}
{\partial\Sigma\over \partial t} = {3\over R}{\partial\over \partial
  R}\left(R^{1/2}{\partial\over {\partial r}}[\nu\Sigma R^{1/2}]\right)
\end{equation}
defines a viscous timescale
\begin{equation}
t_{\rm visc} \sim {R^2\over \nu}
\end{equation}
which can be rewritten using the alpha--prescription as
\begin{equation}
t_{\rm visc} \sim {1\over \alpha}\left({R\over H}\right)^2t_{\rm dyn}
\end{equation}
where $t_{\rm dyn} = (R^3/GM)^{1/2}$ is the local dynamical time.

Thus for example we now know that superhumps -- photometric
modulations in certain CVs, with periods slightly longer than the
binary orbit -- result from the presence of the orbital 3:1 resonance
within a sufficiently large accretion disc (corresponding to a fairly
extreme mass ratio in a CV: cf Whitehurst \& King, 1991; Lubow, 1991,
1992).  Here $t_{\rm visc}$ governs the disc structure, but
variations of it are unimportant for understanding
superhumps. Similarly, the possibility of disc instabilities (see the
talk by Lasota at this meeting) requires one to imagine only two
things: (i) disc structure differs radically if hydrogen is
predominantly ionized or not, and (ii) hotter regions have higher
viscosity and evolve faster than cool ones. Perhaps suprisingly, the
behaviour of the instability in a disc strongly irradiated by the
central (X--ray) source is qualitatively much easier to understand
(King \& Ritter, 1998), since the irradiation traps the disc in the
hot state, allowing a pure viscous--timescale decay.  This is
particularly simple -- a pure exponential -- if the disc is small
enough that the central source keeps it hot at all radii (the
so--called FRED -- fast rise, exponential decay systems).  These cases
and others give straightforward estimates of $\alpha$ as $\alpha \sim
0.1 - 0.4$. 

By now one can arrive at some kind of understanding of how
almost any pattern of light--curve behaviour can be understood using
this picture of disc instability, and possibly allowing for some mild
intrinsic variability (e.g. magnetic spots on the secondary; King \&
Cannizzo, 1998).  In summary we can say that thin--disc `theory' works quite
well provided that we {\it assume} these values of $\alpha$.

However this apparent success comes at a double price. First, it is
entirely ad hoc -- it works (rather like the Old Quantum Theory)
because of a series of fudges and empirical rules. Second, and
more seriously, it means that we are entirely unable to predict, 
or confirm, or deny, {\it global} changes in disc structure, such as 
whether thin disc accretion can make a transition to
advection--dominated (ADAF) flow, or just how and when an accretion
disc creates and powers a jet at its centre. This is analogous to the
inability of pre--nuclear stellar structure theory to predict or
explain supernovae.

\section{The Answer?}

In searching for the true mechanism for angular momentum transport we
need to remember the distinctive feature of accretion discs, that they
simultaneously obey
\begin{equation}
{\partial\over \partial R}(R^2\Omega) > 0,\ \ {\rm and}\ \ 
{\partial\Omega\over \partial R} < 0
\end{equation}
where $\Omega$ is the local angular velocity (the Kepler value
$(GM/R^3)^{1/2}$ in a thin disc). That is, angular momentum increases
outwards, but angular {\it velocity} decreases outwards. The first
property tell us that discs are stable against axisymmetric
perturbations (Rayleigh criterion), so removing a large number of
candidate viscosity mechanisms. Indeed, most purely hydrodynamical mechanisms
are sensitive to the gradient of angular momentum rather than
velocity, and so would if anything transport angular momentum inwards.

The dragging of magnetic fieldlines anchored in an accretion disc 
on the other hand {\it is} sensitive to the angular velocity gradient,
and does offer a promising candidate (Balbus \& Hawley 1991) for the 
transport mechanism. But although this is clear, actually calculating
the full effect of this process is a formidable challenge. In
principle one has to solve the full disc structure self--consistently, 
describing gas motions in full 3D, time--dependent MHD. Most 
theoretical effort so far has gone into trying to use numerical
simulations to quantify the viscous
transport (naively, estimate $\alpha$) in the so--called `shearing
box' approximation. Here one considers a corotating Cartesian box,
plus tidal gravity and Coriolis terms. This is a much more tractable
problem, but has inevitable limitations. Most obviously, the scale of
the box is only $\sim H$, so the simulations are only sensitive to 
high wavenumbers $\sim R/H$, and hence small--scale magnetic fields.  

The results of this procedure are mixed (see King, Pringle \& Livio,
2007 for a recent review).
Fully--ionized shearing--box
simulations tend to give rather small values $\alpha \sim 0.02$,
unless a vertical magnetic field is imposed from the start. Worse,
there is some indication that the value of $\alpha$ is
resolution--dependent (Fromang \& Papaloizou, (2007) in the sense
that $\alpha$ decreases as the numerical resolution of a simulation is
increased. So although
MHD effects are probably the basis of accretion disc viscosity,
current simulations have not conclusively shown this, still less that
the effect is large enough to account for observations. Evidently this
problem will require global disc simulations, and so even more powerful
computers.

\section{Conclusions}

Our current picture of accretion discs is based largely on the
ideas of Shakura and Sunyaev (1973). It
works reasonably
well in a number of areas, provided that we assume 
$\alpha \sim 0.1 - 0.4$. However it is still ad hoc, and is unable to
predict or confirm global changes of disc structure. 
The real basis of accretion disc viscosity is probably magnetic,
as suggested by Balbus and Hawley in 1991. Attempts to
demonstrate this with shearing--box simulations produce viscosities
which are too weak  (corresponding
to $\alpha \la 0.02$) compared with observed constraints
and may even be resolution--dependent.
Balbus and Hawley's paper appeared less than 20 years after Shakura \&
Sunyaev (1973), but 20 years further on still the huge complexity of the
viscosity problem has markedly slowed practical progress. We are still
a long way from a theory of accretion discs with real predictive
power.

\begin{acknowledgements}
I thank many colleagues, particularly Jim Pringle, for insight and 
illumination on these subjects over the years.  My thanks to the
organisers for the superb hospitality and organisation of the
meeting. Research in theoretical astrophysics at Leicester is
supported by an STFC Rolling Grant.
\end{acknowledgements}

\end{document}